\documentclass{aa}  

\bibpunct{(}{)}{;}{a}{}{,} % to follow the A&A style

\usepackage{graphicx}
\usepackage{color}
\usepackage{txfonts}
\usepackage{ulem} % SACARLO ANTES DE ENVIAR: ES PARA TEXTO TACHADO
\usepackage{hyperref}
\definecolor{yaleblue}{rgb}{0.1,0.3,0.9}
\definecolor{bostonuniversityred}{rgb}{0.8, 0.0, 0.0}
\definecolor{lava}{rgb}{0.81, 0.06, 0.13}
\definecolor{forestgreen}{rgb}{0.0, 0.27, 0.13}
\hypersetup{colorlinks=true, linkcolor=lava, urlcolor=forestgreen, citecolor=yaleblue}
\begin{document} 
  % \title{Hydrodinamical model for the Type II SN~2023ixf}
   \title{The progenitor of SN 2023ixf from hydrodynamical modelling}

   \author{M. C. Bersten \inst{1, 2, 3} 
          \and
          M. Orellana \inst{4,5}
          \and
          G. Folatelli\inst{1, 2,3}
          \and
           L. Martinez\inst{1,4}
           \and
          M. P. Piccirilli\inst{2,5}
          \and
          T. Regna\inst{1}
          \and
          L.M. Román~Aguilar\inst{1,2}
          \and
          K. Ertini\inst{1,2}
          %\fnmsep\thanks{Just to show the usage of the elements in the author field}
          }

   \institute{Instituto de Astrof\'isica de La Plata (IALP), CCT-CONICET-UNLP. Paseo del Bosque S/N, B1900FWA, La Plata, Argentina  
       \and Facultad de Ciencias Astronómicas y Geofísicas, Universidad Nacional de La Plata, Paseo del Bosque S/N 1900 La Plata, Buenos Aires, Argentina. 
         \and
         Kavli Institute for the Physics and Mathematics of the Universe (WPI), The University of Tokyo, 5-1-5 Kashiwanoha, Kashiwa,
    Chiba, 277-8583, Japan
         \and
         Universidad Nacional de Río Negro. Sede Andina, Laboratorio de Investigación Científica en Astronomía, Anasagasti 1463, Bariloche (8400), Argentina
         \and
         Consejo Nacional de Investigaciones Científicas y Técnicas (CONICET),  Godoy Cruz 2290, 1425 Ciudad Aut\'onoma de Buenos Aires, Argentina. 
         }
   \offprints{M. Bersten\\
   \email{mbersten@fcaglp.unlp.edu.ar}}
%   \email{c.ptolemy@hipparch.uheaven.space}
%   \thanks{The university of heaven temporarily does not accept e-mails}
%             }
\titlerunning{Hydrodinamical model of SN 2023ixf}
\authorrunning{Bersten, M. et al.}
   %\date{Received xx; accepted xx}
   \date{Submitted 6-10-2023}

% \abstract{}{}{}{}{} 
% 5 {} token are mandatory
 
  \abstract
   {Supernova (SN) 2023ixf is among the most nearby Type II SNe in the last decades. As such, there is a wealth of observational data of both the event itself and of the associated object identified in pre-explosion images. This allows to perform a variety of studies that aim at determining the SN properties and the nature of the putative progenitor star.  Modelling of the light curve is a powerful method to derive physical properties independently of direct progenitor analyzes.}
   {To investigate the physical nature of SN~2023ixf based on hydrodynamical modelling of its bolometric light curve and expansion velocities during the complete photospheric phase.}
   {A grid of one dimensional explosions was calculated for evolved stars of different masses. 
   We derived properties of SN~2023ixf and its progenitor by comparing our models with the observations.}
   {The observations at $t \gtrsim 20$ days are well reproduced by the explosion of a star with zero age main sequence mass of $M_\mathrm{ZAMS} = 12 \, M_\odot$, an explosion energy of $1.2 \times 10^{51}$ erg, and a nickel production of $0.05 \, M_\odot$. This indicates that SN~2023ixf was a normal event. Our modelling suggests a limit of $M_\mathrm{ZAMS} < 15 \, M_\odot$ and therefore favours the low mass range among the results from pre-explosion observations.}
    {}

   \maketitle
%-------------------------------------------------------------------
\section{Introduction}
\label{sec:intro}
Supernova (SN) 2023ixf was discovered in 2023 May 19 17:27:15.00 UT in the galaxy M101 \citep{itagaki23_tns} and it was classified as a Type II SN (SN II) \citep[][]{perley+23_tns,biancardi23}. This object is among the nearest core collapse SNe (CC-SNe) observed in recent years. Due to its proximity, it has attracted the attention of the entire community and it triggered extensive observations by professional and amateur astronomers alike. Optical, near infrared (IR) and ultraviolet (UV) follow-up observations started within one day from the explosion. Early spectroscopy showed flash-ionization emission features lasting for several days, which is indicative of the presence of a dense circumstellar material (CSM) \citep[][]{sutaria+23_atel,perley+23_tns,benzvi+23_tns,stritzinger+23_tns,smith+23,bostroem+23,yamanaka+23,teja+23,jacobson-galan+23,hiramatsu+23}. This was further supported by X-ray \citep{mereminskiy+23,chandra+23,grefenstette+23,panjkov+23}, radio \citep{matthews+23}, and polarimetry \citep{vasylev+23} observations. 
%Although no significant neutrino signal was detected by any of the neutrino experiments \citep{nus}, prompt observations  extensively covered most of the electromagnetic spectrum.

The site of SN~2023ixf had been observed with several facilities during years before the explosion, particularly with the \textit{Hubble Space Telescope} (\textit{HST}) in the optical and the \textit{Spitzer Space Telescope} in the IR. Various studies have been published to date that analyze the pre-SN photometry and derive properties of the putative progenitor object, most importantly its initial mass. Although all works agree on the identification of the progenitor candidate as a dust-obscured red supergiant (RSG) star, there are discrepancies on the derived zero-age main sequence mass ($M_\mathrm{ZAMS}$). From spectral energy distribution fits including an RSG spectrum plus thermal emission from dust, and comparison with stellar evolution tracks, several authors found that the pre-SN object was compatible with a mass of $M_\mathrm{ZAMS} = 10 - 15 \, M_\odot$ \citep{neustadt+23,kilpatrick+23,vandyk+23b,xiang+23}. Similar analyzes as above provided higher initial masses of $M_\mathrm{ZAMS} \approx 16 - 18 \, M_\odot$ due to the derivation of a higher progenitor luminosity \citep{jencson+23,niu+23,Qin+23}. On the other hand, \citet{pledger+23} estimated a slightly smaller mass of $M_\mathrm{ZAMS} = 8 - 10 M_\odot$, although solely based on the \textit{HST} images. From an environmental study of the SN site \citet{niu+23} estimated the youngest stellar population to be $\approx 12$ Myr old and thus suggested a progenitor mass of $M_\mathrm{ZAMS} = 17 - 19 \, M_\odot$. Finally, \citet{soraisam+23} analyzed the IR variability of the progenitor candidate and derived its luminosity from a pulsational period-luminosity relation, which allowed them to obtain a distance and extinction-independent mass of $M_\mathrm{ZAMS} = 20 \pm 4 \, M_\odot$. 

Given the wide range of progenitor mass estimates obtained from the pre-SN data, it is crucial to contrast those results by using alternative methods. One such method is the hydrodynamical modeling compared with the SN bolometric light curve and expansion velocity evolution. The present work is the first attempt of such an analysis using observations of SN~2023ixf throughout the plateau phase and on to the radioactive tail phase. This allows us to derive progenitor properties in an independent manner from those of pre-SN studies. Section~\ref{sec:Lbol} presents the data and the calculation of bolometric luminosities and spectral line velocities. The hydrodynamical modelling is described in Sect.~\ref{sec:hydro}. Finally, in Sect.~\ref{sec:final} we summarize our results and compare the derived progenitor properties with those of previous works.

%--------------------------------------------------------------------
\section{Bolometric light curve and expansion velocities}
\label{sec:Lbol}
In order to calculate the observed bolometric light curve (LC) for SN~2023ixf we used public photometry available in the $B$ and $V$ bands from the American Association of Variable Star Observers (AAVSO) web page\footnote{www.aavso.org}. The AAVSO server provides a compilation of photometric measurements from different observers around the world. More than 2000 data points were available in the $B$ band, and over 6000 points in the $V$ band, in both cases covering over 100 days of the SN evolution. We adopted the mean magnitudes computed in bins of 1 day after rejecting discrepant observations. The dispersion of points within each bin was always below $0.1$ mag. Intrinsic $(B-V)$ colors were computed using Milky-Way and host-galaxy color-excesses of $E(B-V)_\mathrm{MW}=0.008$ mag \citep{schlafly+11} and $E(B-V)_\mathrm{host}=0.031$ mag \citep{lundquist+23_tns}, respectively. We then used the $(B-V)$ color-based bolometric corrections as calibrated by \citet[][]{martinez22a} to derive bolometric magnitudes. Finally, bolometric luminosities were computed by adopting a distance to M101 of $6.85\pm 0.15$ Mpc \citep{riess+22}. The resulting bolometric LC is shown in Fig.~\ref{model}. We compute the rest-frame time relative to the explosion time of $\mathrm{MJD} = 60082.75$ given by \citet{hosseinzadeh+23}, and adopting a redshift of $z = 0.0008$ from the NASA/IPAC Extragalactic Database (NED).

Given the exceptional wavelength coverage and temporal sampling of SN~2023ixf at early times \citet{martinez+23} were able to compute a detailed bolometric LC until 19 days after explosion. They performed the calculations via integration of the spectral energy distributions and black-body extrapolations toward shorter and longer wavelengths. For comparison, we show this LC with gray points in Fig.~\ref{model}. We note that after $\approx$5 days since explosion, both bolometric LCs agree fairly well with each other. This suggests that the complete bolometric LC presented here can be reliably used to derive overall physical parameters of SN~2023ixf as we do in Sect.~\ref{sec:hydro}.

The hydrodynamical modelling  can be additionally constrained by using  an estimate of the velocity at the photosphere as it evolves with time. In order to estimate this photospheric velocity we used public spectra from the Weizmann Interactive Supernova Data Repository (WISeREP)\footnote{\url{https://wiserep.weizmann.ac.il}} (\citealp[]{wiserep}), selecting those where the Fe~{\sc ii}\,$\lambda$5169 line could be identified (which occurred after $\approx 25$ days from the explosion). This criterion led us to use three spectra from the Dark Energy Spectroscopic Instrument (DESI) (\citealt{levi19}) at the 4m Mayall Telescope at Kitt Peak National Observatory and one spectrum uploaded by the Transient Name Server (TNS\footnote{\url{https://www.wis-tns.org/}}) without information about the telescope and instrument. We measured the wavelength at the absorption minimum of the spectral lines and thereby we computed line velocities from the Doppler shifts relative to the rest wavelength of those lines.
We performed this for the H$\alpha$, H$\beta$ and the Fe~{\sc ii}\,$\lambda$5169 lines, which are fairly uncontaminated by other absorptions and can be identified and measured through most of the plateau phase. The resulting velocities are plotted in Fig.~\ref{model}. We note that the Fe~{\sc ii} velocities are systematically lower than those from H$\alpha$ and H$\beta$. %Quedaron las 2, mencionamos Hbeta. Veamos porque ademas hay que ver como quedan las velos photospericas de Laureano y ahi decidimos que hacer.
This is usually the case in SNe and it is due to the fact that the weaker Fe~{\sc ii} absorption is formed deeper in the SN ejecta. This in turn justifies its use as a better indicator than H$\alpha$ for the photospheric velocity \citep{dessart+05}. 

\section{Hydrodynamical modelling}
\label{sec:hydro}
To derive physical parameters for SN~2023ixf we compare the bolometric LC and the expansion velocities derived in Sect.~\ref{sec:Lbol} with a grid of explosion models. 
The models are computed using the one-dimensional Lagrangian local thermodynamic equilibrium (LTE) radiation hydrodynamics code presented by \citet{bersten+11}.  

As initial conditions (or pre-SN models) we adopted hydrostatic structures calculated using the publicly available stellar evolution code MESA6 version 22.6. \citep{paxton+11,paxton+13,paxton+15,paxton+18,paxton+19, jermyn+23}. Specifically, we produced models with zero-age main sequence  (ZAMS) masses of 12, 15, 20, and 22 M$_\odot$ for which we followed the complete evolution of the star from ZAMS to the pre-collapse condition\footnote{Defined as the time when any location inside the iron core reaches an infall velocity of 1000 km~s$^{-1}$}. These models were computed assuming no rotation, and a solar metallicity \citep[$Z = 0.0142$; see][for more details on the physical assumptions]{martinez+23}.
%\textbf{ In addition, two different set of models were computed, one assuming an standard wind for massive stars \citep{vink+01,dejager+88} and other where the the mass loss was reduced by a factor 3 assuming the cumpling effect \citep{Smith14b,Humphreys20}} The different wind assumptions allowed us to have different pre-SN conditions, mass and radii, at the moment of the explosion for a given  M$_\mathrm{ZAMS}$.

It is known that the evolutionary models alone fail to reproduce the early emission ($t \lesssim 20$ days) observed in many SNe~II. An ad hoc modification of the outermost layers of the star is usually done to account for the existence of a possible nearby CSM ejected by the star during its evolution prior to the explosion \citep{moriya+11,morozova+18}. %by a mechanism that is not entirely clear \citep[see][for some of the proposals] {quataert+12,suarez-madrigal+13,smith+14,fuller17}. 
Although the focus of this Letter is to analyze the bolometric light curve of SN~2023ixf at times when the effect of the CSM is no longer dominant, we do include in our pre-SN models the presence of a steady-state wind attached to the stellar structure. We do this by modifying the initial density profile assuming an external density distribution with a radial dependence like $\rho \propto r^{-2}$. The mass ($M_\mathrm{CSM}$) and extension ($R_\mathrm{CSM}$) of the CSM are free parameters that can be inferred from the modelling of the early data. Nevertheless, we note that the values of $M_\mathrm{CSM}$ and $R_\mathrm{CSM}$ are not univocal and they may also depend on the assumed density and velocity distribution of the wind. A detailed analysis of the CSM properties is presented in our companion paper \citep[see][]{martinez+23}. 
Here we simply assume a steady wind with a constant velocity of 10 km~s$^{-1}$ as typically adopted for RSG stars.

\begin{figure}
   \resizebox{\hsize}{!}{\includegraphics{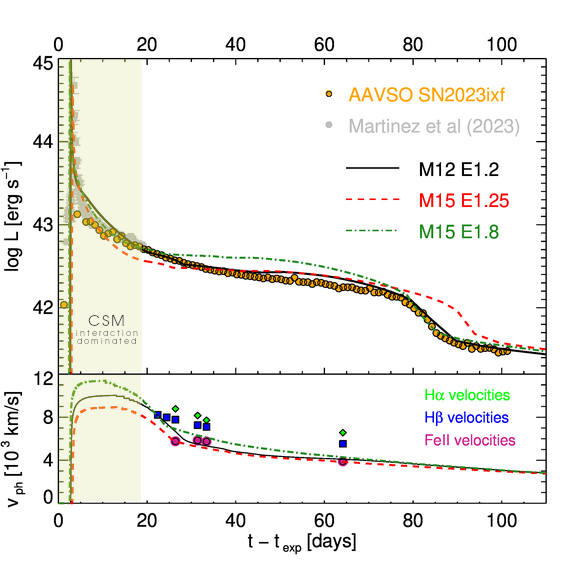}}
%   \centering
%   \includegraphics[scale=0.6]{mejores_5.png}
   \caption{Hydrodynamical models (lines) compared with observations of SN~2023ixf (points). \textit{Upper panel:} bolometric light curve; \textit{lower panel:} expansion velocities. We have shadowed the approximate time range when the emission is dominated by CSM interaction.
   (i.e. $t < 20$ d). This is based on previous results from the literature \citep{morozova+18, martinez+23}.
%   The time for the end of this phase has not a precise definition, our choice is based on previous results from the literature.
   The lower mass model of $M_\mathrm{ZAMS} = 12 M_\odot$ produces a better match to the observations than the $15 M_\odot$ model.  
 Particularly,   the higher-mass model  with $1.25 \times 10^{51}$ erg produces a longer plateau duration than what is observed while the model with an energy of $1.8 \times 10^{51}$ erg provided the correct plateau duration but overestimated the observed plateau luminosity. These issues cannot be solved by modifying other parameters (see discussion in Sect.~\ref{sec:hydro}). This suggests that $M_\mathrm{ZAMS} < 15 M_\odot$.}
   \label{model} %
   \end{figure}

The focus of this work is to derive global SN parameters such as the ejecta mass, explosion energy and nickel production, from the modelling of the LC and velocities during the plateau and radioactive tail phases (i.e. at $t \gtrsim 20$ days). This analysis can be decoupled from that of the CSM interaction dominated era \citep[see]{morozova+18,hillier+19,martinez+22b} that is shown as a shaded region in Fig.~\ref{model}.
%To derive physical parameter through the hydrodynamical modelling it is possible to divide the problem into two parts (see \cite{morozova+18,hillier+19,martinez+22b} ): 1) the early modelling  ( $t \lesssim 20$ days) dominated by the properties of the CSM (see above discussion)  and 2) the plateau and radioactive tail modelling. Global properties of progenitor and explosion as the ejecta mass, explosion energy and nickel production are derived modelling  the LC and velocities in the second stage. 
The sensitivity of the global parameters on the observed data has been studied in numerous works \citep[see e.g.,][]{Utrobin07,bersten+11}. 
%And this is the focus of this work. For clarity  we  have shaded the CSM dominated region in Figure~\ref{model} which is not deeply analyzed here.

Despite having a grid of models with a wide range of $M_\mathrm{ZAMS}$, from an initial inspection we noted that only models with pre-SN masses constrained to $\lesssim 15$ M$_\odot$ were able to reproduce the observations. This was due to the relatively short plateau duration and high luminosity (see Sect.~\ref{sec:Lbol}), which disfavoured more massive pre-SN configurations.
This is also based on our general knowledge of how the explosion models behave when physical parameters vary \citep[see e.g.,][]{Utrobin07,bersten+11}.  Therefore only models with $M_\mathrm{ZAMS}$ of 12 and 15 $M_\odot$ were more deeply explored.  SN explosions were simulated from these initial models and adopting different explosion energies ($E_\mathrm{exp}$), nickel masses ($M_\mathrm{^{56}Ni}$), and nickel distribution. Our preferred model is presented with a solid line in Fig.~\ref{model}, and it corresponds to $M_\mathrm{ZAMS} =  12 M_\odot$, $E_\mathrm{exp} = 1.2 \times 10^{51}$ erg, $M_\mathrm{^{56}Ni} = 0.05 M_\odot$, with an almost complete mixing of the radioactive material within the ejecta. This model has a pre-SN mass of $10.9 M_\odot$, a radius of $720 R_\odot$. The innermost $1.5 M_\odot$ of the pre-SN structure is assumed to collapse into a compact remnant. 

For comparison in Fig.~\ref{model} we present models based on a more massive progenitor with $M_\mathrm{ZAMS} = 15 M_\odot$ (dashed and dot-dashed lines) which have a pre-SN mass and radius of $12.7 M_\odot$ and $970 R_\odot$ respectively.  For this model (denoted by M15) we show the calculations for two explosion energies of $1.25 \times 10^{51}$ erg and $1.8 \times 10^{51}$ erg assuming a remnant mass of  $1.8 M_\odot$ and a nickel mass of $M_\mathrm{^{56}Ni} = 0.05 M_\odot$. From the figure it is clear that the model with lower energy (M15 E1.25) is able to reproduce the plateau luminosity but it overestimates the plateau duration. On the other hand, the larger energy model (M15 E1.8) provides the right plateau duration but overestimates the plateau luminosity.
 %From Figure~\ref{model} it is clear that this more mas
%sive model produces a longer plateau duration than what is observed. This cannot be reduced by increasing the explosion energy because this would lead to a more luminous plateau in contrast with the observations.
The other parameter that can have an effect on the plateau duration (and its shape), although much weaker than the expected effect from pre-SN mass and explosion energy, is the nickel mixing. We tested the effect of varying the nickel mixing but we did not find an improvement compared with the presented model. 

Although our main goal did not involve the modelling of the early evolution, for completeness here we provide the adopted CSM parameters for the models presented in Fig.~\ref{model}. These are: $M_\mathrm{CSM} = 0.4 M_\odot$, and $R_\mathrm{CSM} = 2000 R_\odot$. These values correspond to a mass loss rate of $0.14 M_\odot$~yr$^{-1}$ under the hypothesis of a steady wind. We note, however, that the match to the observations is poor at times $\lesssim$ 10 days.  A detailed analysis and modelling of the early evolution of SN~2023ixf and the wind properties required to reproduce the maximum luminosity and its timescale are presented in \citet{martinez+23}. Nevertheless, our conclusions remain unchanged about the main physical parameters that reproduce the overall SN evolution.

We found that the model that better reproduces the observations of SN~2023ixf is the one with the lowest pre-SN mass available in our grid. Although in principle we cannot rule out less massive progenitors, we note that the initial mass of our preferred model ($M_\mathrm{ZAMS} = 12 M_\odot$) and our constraint of $M_\mathrm{ZAMS} < 15 M_\odot$ favours the lower range of progenitor masses derived in the literature from studies of the pre-SN observations (see Sect.~\ref{sec:intro}).
%

%-------------------------------------- Two column figure (place early!)

\noindent

%--------------------------------------------------- One column table
   \begin{table}
      \caption[]{Bolometric light-curve parameters as defined by \citet{martinez22a}. Average and dispersion values are given from the CSP-I sample of SNe~II for comparison (see text).
      } 
         \label{tab:lcpars}
      \centering
          \begin{tabular}{lcc}
         %\begin{array}{p{0.5\linewidth}l}
            \hline
            \noalign{\smallskip}
            Parameter      &  SN~2023ixf   & CSP-I \\
            \noalign{\smallskip}
            \hline
            \noalign{\smallskip}
            $M_\mathrm{bol,end}$ (mag)      & $-17.18 (0.06)$ & $-16.2 (0.6)$\\
            $M_\mathrm{bol,tail}$ (mag)     & $-14.77 (0.04)$ & $-14.8 (0.3)$\\
            $s_1$ (mag / 100 d)&  $5.53 (0.91)$ & $4.59 (2.84)$\\
            $s_2$ (mag / 100 d)&  $1.84 (0.56)$ & $0.81 (0.91)$\\
            $s_3$ (mag / 100 d)&  $1.71 (0.74)$ & $1.38 (0.62)$\\
            $C_d$ (d)             &  $29.66 (5.31)$ & $26.9 (4.3)$\\
            $pd$ (d)              &  $53.42 (5.23)$ & $75.0 (26.2)$\\
            $optd$ (d)            &  $83.08 (0.08)$ & $104.3 (19.3)$\\
            \noalign{\smallskip}
            \hline
        % \end{array}   
            
          \end{tabular}
              
   \end{table}
%
   
%
%                                                One column figu
%-----------------------------------------------------------------
    
\section{Conclusions}
\label{sec:final}
We present the first hydrodynamical modelling of the bolometric LC and photospheric velocity evolution of SN~2023ixf along the complete extent of the plateau phase and the onset of the radioactive tail. This allows us to obtain overall physical parameters for this SN and its progenitor. Our results suggest that SN~2023ixf originated from the explosion of a $12 M_\odot$ (ZAMS) mass star with an explosion energy of $1.2 \times 10^{51}$ erg, and a $\mathrm{^{56}Ni}$ production of $0.05 M_\odot$. The exploded RSG star had a mass of $10.9 M_\odot$, and a radius of $720 R_\odot$ at the final stage of its evolution.
This indicates that SN~2023ixf was a normal Type~II event as it is also concluded from our comparison of LC morphological parameters with a large sample of SNe~II \citep{martinez22a,martinez+22b}.

The model parameters above reproduce the overall shape of the LC starting after $\approx 10$ days since the explosion. At earlier times, some extra emission is required to match the observations. As suggested in previous works, this extra flux can arise from the interaction between the SN ejecta and some pre-existing CSM. We include such an effect in our calculations although a definitive study of the CSM interaction is left to a separate work \citep{martinez+23}. Our conclusions about the main SN properties are not affected by a possible change in the CSM configuration.

Numerous studies have analyzed the pre-explosion observations of the SN site.  There is a consensus on the progenitor identification as a dusty RSG star. However, a wide range of $M_\mathrm{ZAMS}$ from $\approx 10$ to over $20 M_\odot$ were derived by different authors (see Sect.~\ref{sec:intro}). Our hydrodinamical modelling provides an independent mass estimate and therefore can help to discriminate among the proposed masses. Our analysis suggests that the progenitor of SN~2023ixf was an RSG star with $M_\mathrm{ZAMS} < 15 M_\odot$. This is in line with the relatively low masses estimated from pre-SN spectral energy distribution (SED) fits by  \citet{neustadt+23,kilpatrick+23,vandyk+23b} and \citet{xiang+23}, and marginally in agreement with the result by \citet{jencson+23}. Higher masses are disfavored, such as those obtained also from SED fits by \citet{niu+23,Qin+23}, from environmental studies by \citet{niu+23}, and from IR variability by \citet{soraisam+23}. Future observations such as revisiting the SN site to verify the disappearance of the progenitor candidate, or obtaining late-time spectroscopy during the nebular phase will be necessary to further understand the nature of SN~2023ixf.  

%Next, we expect SN2023ixf to deserve other kind of studies, and hopefully to reach a consistent scenario with the posterior radio analysis as in the case of other nearby type II SNe \citep{2018nayana}.

\begin{acknowledgements}
      We gratefully acknowledge the variable star observations from the AAVSO International Database,
      contributed by observers worldwide and used in this research.
      M.O. acknowledges support from UNRN PI2022 40B1039 grant. L.M. acknowledges support from a CONICET fellowship and  UNRN~PI2022~40B1039 grant.
\end{acknowledgements}

%\facility{AAVSO}


\bibliographystyle{aa} % style aa.bst
\bibliography{SN2023ixf} % references in SN2023ixf.bib

%\begin{thebibliography}{}

%  \bibitem[Baker(1966)]{baker} Baker, N. 1966,
%      in Stellar Evolution,
%      ed.\ R. F. Stein,\& A. G. W. Cameron
%      (Plenum, New York) 333

%   \bibitem[Balluch(1988)]{balluch} Balluch, M. 1988,
%      A\&A, 200, 58

%  \end{thebibliography}

\end{document}